# Excitation of a Si/SiGe quantum dot using an on-chip microwave antenna

E. Kawakami[1], P. Scarlino[1], L. R. Schreiber[1,3], J. R. Prance[2], D. E. Savage[2], M. G. Lagally[2], M. A. Eriksson[2], and L. M. K. Vandersypen[1]

[1]Kavli Institute of Nanoscience, TU Delft, Lorentzweg 1, 2628 CJ Delft, The Netherlands

[2]University of Wisconsin-Madison, Madison, WI 53706

[3]II. Institute of Physics, RWTH Aachen University, 52074 Aachen, Germany

We report transport measurements on a Si/SiGe quantum dot subject to microwave excitation via an on-chip antenna. The response shows signatures of photon-assisted tunneling and only a small effect on charge stability. We also explore the use of a d.c. current applied to the antenna for generating tunable, local magnetic field gradients and put bounds on the achievable field gradients, limited by heating of the reservoirs.

Electron spins in gate-defined semiconductor quantum dots are promising candidates for quantum bits because of their high tunability and scalability. One and two qubit manipulation was experimentally demonstrated in GaAs quantum dots[1,2,3,4,5]. Recently, it has been reported that the electron spin dephasing time, measured via two-spin coherent exchange, is ~40 times longer in Si/SiGe quantum dots than in GaAs dots[6]. This arises from the much weaker hyperfine coupling in silicon. By using predominantly nuclear-spin free $^{28}$Si, the hyperfine coupling is further reduced and dephasing times are considerably longer[7]. Spin-orbit mediated spin relaxation is also slow, with typical timescales upwards of 1 ms[8]. This indicates the strong potential of Si/SiGe quantum dots in quantum information processing.

Achieving single-spin manipulation would be an important milestone in this material system. In GaAs, single-spin rotations were first realized using electron spin resonance (ESR), by applying a local oscillating magnetic field on resonance with the electron spin Larmor precession in a large static magnetic field[5]. The oscillating magnetic field was generated by applying an a.c. excitation to a microwave antenna located next to the quantum dots. Single-spin manipulation has also been realized via oscillating electric fields, which can couple to the spins via spin-orbit interaction[9] or a local magnetic field gradient from a micromagnet[10]. Exploiting spin-orbit interaction is experimentally the simplest approach as it does not require a local micromagnet or antenna. However, spin-orbit coupling is too weak in Si/SiGe, and so one of the other methods is required for coherent single-spin control.

In single-spin resonance experiments, the spin state can be conveniently detected in a double quantum dot tuned to the so-called Pauli spin blockade regime[11]. In this regime, two-electron spin singlets are distinguished from spin triplets. In GaAs, the effective magnetic field gradient created by the nuclear spins quickly mixes the $m = 0$ triplet $T^0$ with the singlet S so that spin blockade differentiates between parallel and anti-parallel spins. This differentiation is an essential ingredient for the detection of ESR using Pauli blockade[4,5,9]. In Si/SiGe quantum dots, the nuclear fields are much weaker and detection of ESR may fail unless a magnetic field gradient is created by other means.

Here we use an on-chip antenna, adjacent to a Si/SiGe quantum dot, to produce microwave excitation and static magnetic field gradients, by driving with both d.c. and a.c. currents. We study the resulting effects on the transport characteristics of a single quantum dot and evaluate the potential for using such an antenna to perform single-spin ESR experiments in Si/SiGe devices.

Our device is fabricated on a phosphorus-doped $Si/Si_{0.7}Ge_{0.3}$ heterostructure with a strained Si quantum well approximately 75 nm below the surface. Palladium surface gates labelled 1-9 in Figure 1(a) can be used to form a single dot or a double dot. The experiments shown here use a single dot. An on-chip antenna (Ti/Au, 5 nm/305 nm) is fabricated close to the dot gates, as shown in Figure 1(a). D.c. and a.c. current through the antenna are combined via a bias-T placed at the 1 K stage of the refrigerator. All gates are connected to room temperature voltage sources via RC and copper powder filters mounted below the mixing chamber and room temperature pi-filters. Both ends of the antenna are connected to high-frequency lines.

First, we test whether the charge stability of the Si/SiGe quantum dot in the few electron regime is affected by microwave excitation of the antenna. The electric field component of the excitation may perturb and rearrange charges trapped in the substrate, thereby generating electrical noise. An a.c. excitation of $f = 20\,\text{GHz}$ is applied to both ends of the antenna. The ratio of the microwave electric versus magnetic field strength at the dot depends on the relative phase of the excitation at the two ends of the antenna. In the measurements reported here the relative phase is arbitrary.

To probe charge stability, we repeatedly measure Coulomb peaks in the left dot by sweeping gate 4 with and without microwave radiation applied to the antenna. The measured Coulomb peaks from 29 repetitions are superimposed in Figure 1(b). The microwave excitation broadens the Coulomb peaks to the point where they begin to overlap. The broadening is due to heating and photon-assisted tunnelling, which is discussed further below. The microwave electric field applied here is the upper limit of what could be applied in ESR experiments.

We evaluate $\sigma_{V_{gate4}}$ charge stability in units of gate voltage, as charge noise in the substrate affects the dot in the same way as gate voltage noise. We calculate $\sigma_{V_{gate4}}$ over $-483 \text{ mV} < V_{gate4} < -300 \text{ mV}$, restricting ourselves to current levels $I_{dot} < -2$ pA, where $\sigma_{V_{gate4}}$ is $\sigma_{I_{dot}}$, the standard deviation of the current level over 29 repetitions, divided by $dI_{dot}/dV_{gate4}$, the numerical derivative of the current with respect to gate voltage ($\sigma_{V_{gate4}} = \sigma_{I_{dot}} / (dI_{dot}/dV_{gate4})$). The histogram of $\sigma_{V_{gate4}}$ is shown in Figure 1(c). We see that microwave excitation produces only a small shift in the distribution of $\sigma_{V_{gate4}}$, i.e., there is only a small increase in charge noise, even with a high power applied to the antenna.

We now analyze the Coulomb peak shape in the presence of microwave power. We show the response for different excitation frequencies and microwave powers in Figure 2(a). The traces show the typical characteristics of photon-assisted tunnelling (PAT) in a single quantum dot when the microwave field couples asymmetrically to the device[12,13,14,15]. Specifically, the microwaves can couple differently to the dot, to the source, and to the drain, as discussed in Ref. 13. This results in unequal voltage drops at the left and right tunnel barriers due to the a.c. excitation. Figure 2(b,c) and Figure 2(d,e) depict the extreme cases, where there is an a.c. voltage drop only across the right and left barrier, respectively. When the dot level is above the Fermi level of a reservoir by exactly the microwave energy, tunnelling from the reservoir into the dot across this barrier is made possible through photon-assisted tunnelling (PAT), as depicted by the long red arrows in Figures 2(c,e). Similarly, the dot can be depopulated by PAT if it is below the Fermi level of a reservoir by exactly the microwave energy, as shown in Figure 2(b,d). Once the dot is populated (depopulated) through PAT, it can depopulate (populate) by tunnelling through either barrier, as represented by the short gray arrows in Figures 2(b-e). The sequence of population and depopulation induces a non-zero net electron flow as indicated by the blue arrows at the bottom of Figures 2(b-e), which would be present even without a voltage bias across the dot[16,17]. The pumping contribution, which is asymmetric in gate voltage, adds to the gate-voltage symmetric contribution from the bias. A further asymmetric contribution to net current can arise from tunnel-barrier modulation as discussed in Refs. 12, 18 and 19.

The asymmetry of the Coulomb peaks for $f = 13.5\,\text{GHz}$ and $16.5\,\text{GHz}$ in Figure 2(a) indicates that the left barrier has the larger a.c. voltage drop, corresponding to the situation of Figures 2(d,e). On the right side of the peaks (shown in Figure 2(d)) PAT leads to extra negative current and on the left side (the case of Figure 2(e)) to extra positive current. Thus the single dot operates as an electron pump under microwave irradiation. As expected, the pumping current becomes more pronounced with stronger microwave power[12,13], and eventually it can dominate transport through the dot. The asymmetry of the peaks is reversed for $f = 20\,\text{GHz}$, indicating that here the right barrier has the larger a.c. voltage drop, corresponding to the situation of Figures 2(b,c). A qualitatively similar frequency and power dependence of the Coulomb peak shape was observed when applying microwave excitation to gate 2, indicating that these observations are not specific to excitation of the antenna.

Next we turn to applying a d.c. current to the antenna, creating a local static magnetic field gradient at the position of the dots. To detect ESR using transport measurements in the spin blockade regime (often the method of choice[4,5,9]), S-T$^0$ mixing, which lifts spin blockade, should be faster than 1 MHz. This gives current levels ~ 160 fA, which is a good target value to give observable contrast between parallel and anti-parallel spins[5]. Based on numerical simulations of the magnetic field profile generated by the antenna, we estimate that a 4 mA d.c. current produces a ~ 40 $\mu$T field difference between two dots that are 30 nm apart and separated from the antenna by 200 nm (the lateral distance between the center of the two dots and the end of the on-chip antenna). A ~ 40 $\mu$T field difference is ~3 times higher than the intrinsic nuclear field difference in Si/SiGe[6], and would give a 1.1 MHz S-T$^0$ mixing rate for a *g*-factor of 2. A further contribution to singlet-triplet mixing arises when the microwave field amplitude is different in the two dots (with this sample, we expect 10% of amplitude difference), causing the spins to rotate at different Rabi frequencies[5]. In this sample, we expect 10% of amplitude difference.

The d.c. current that can be applied is ultimately limited by Joule heating. This increases the temperature of the reservoirs and broadens the Coulomb blockade peaks. We have determined the heating of the electron reservoirs by the d.c. biased antenna. Figures 3(a-c) show a Coulomb peak measured while applying d.c. currents up to 6 mA for three different source-drain voltages $V_{SD} = -58\ \mu V$, $-8\ \mu V$ and $42\ \mu V$. The horizontal axis of Figures 3(a-c) is the electrochemical potential of the dot, converted from $V_{gate2}$ using the conversion factor 50 $\mu$eV/mV (extracted from Coulomb diamonds).

According to the Landauer formula[20,21,22] the current through a single quantum dot is given by

$$I = -\frac{2e}{h} \int_{-\infty}^{\infty} (f_S(\varepsilon) - f_D(\varepsilon))\tau(\varepsilon)d\varepsilon, \quad (1)$$

where $\tau(\varepsilon)$ is the transmission coefficient of the quantum dot as a function of energy $\varepsilon$,

$$\tau(\varepsilon) = \frac{(\Gamma/2)\pi}{(\Gamma/2)^2 + \varepsilon^2}, \quad (2)$$

and $f_S(\varepsilon)$ ($f_D(\varepsilon)$) is the Fermi distribution function of the source (drain)

$$f_i(\varepsilon) = \exp\left(\frac{\varepsilon - \mu_i}{kT_i}\right) + 1 \quad (i = S, D) \quad (3)$$

with $\mu_S - \mu_D = V_{SD}$, $k$ Boltzmann's constant and $T$ the temperature. If the tunnel coupling $\Gamma$ between the dot and the reservoir is much less than $kT$, the transmission coefficient is well approximated by a delta function[23]: $\tau(\varepsilon) \approx \delta(\varepsilon)$. Fitting the Coulomb peak for $V_{SD} = -8\mu V$, without any d.c. current or a.c. excitation, (* symbols in Figure 3 (b)) with Eq.(1) gives $\Gamma \sim 0.9\ \mu eV$ ($\sim 10$ mK). Therefore we can apply the delta function approximation and the current can be rewritten as

$$I = -\frac{2e}{h}(f_S(\varepsilon) - f_D(\varepsilon)). \quad (4)$$

The Coulomb peaks of the right dot with $I_{DC} = 0$ mA, 1 mA, 2 mA and 3 mA, and for $V_{SD} = -58\mu V$, $-8\ \mu V$ and $42\ \mu V$, are fitted to Eq.(4) (solid lines in Figures 3(a-c)). This expression applies as long as transport occurs via a single quantum dot level only, i.e. when the energy level spacing is larger than the temperature of the reservoirs. The smallest energy splitting in Si/SiGe quantum dots is usually the valley-orbit splitting, which is typically of the order of 100 $\mu eV$ to 300 $\mu eV$[24,25]. Thus we assume that Eq.(4) is a good fitting model below 1.2 K ($\sim 100\ \mu eV$). Figure 3 (d) shows the temperatures in the source and drain reservoirs ($T_S$ and $T_D$) obtained from the fits. As expected, both temperatures increase with the applied d.c. current, and the temperature in the source reservoir, which is closest to the on-chip antenna, is higher in all cases. The arrows in Figure 3 (a-c) show the direction of the electron flow at different gate voltages. At certain points, the difference of the temperatures in the two reservoirs can induce electron flow in the opposite direction of the applied bias.

This looks superficially similar to the pumping currents due to PAT. We can directly compare the Coulomb peaks in Figure 3(a) with the Coulomb peak at $V_{gate2} \sim -355$ mV in Figure 2(a), since they are measured in the same configuration. We see that at high microwave power, the Coulomb peak shape in Figure 2(a) for the case of 13.5 GHz and 16.5 GHz has an opposite asymmetry to the Coulomb peak seen in Figure 3(a), which is caused by heating below the antenna. We take this as evidence that at high power, photon-assisted tunnelling effects are dominant over heating via phonons.

We note that an asymmetric Coulomb peak is observed for $V_{SD} \sim 0$ even without d.c. or a.c. current (see Figure 3(b)). From fits to Eq.(4), we find that the temperature difference between the two reservoirs is around 100 mK even for $I_{DC} = 0$ as shown in Figure 3(d). The d.c. line connected to the on-chip antenna goes to the room temperature current source without filtering. Thus Johnson–Nyquist noise coming from the room temperature may cause heating beneath the antenna, giving a temperature difference between two reservoirs. Similar asymmetric heating of the reservoirs was found when measuring the left dot, and when the constriction between gates 1 and 9 was pinched off.

In spin qubit measurements, the temperature should be smaller than the energy scale that is important for initialization and single-shot read-out[2,26]: the Zeeman energy, which is ~ 100 μeV/T in silicon. Another relevant energy scale is the lowest orbital splitting, or the valley-orbit splitting, typically at least 100 µeV. Other energy scales such as the charging energy are significantly larger. The temperature should therefore remain well below ~ 1 K, and from the results shown in Figure 3(d), this implies that we should limit the d.c. current to 2 mA. This is about two times less than the 4 mA needed for efficient detection of ESR-induced Rabi oscillations, as discussed above (We note that an oscillation can be detected even without a gradient if the magnetic excitation differs in strength between the two dots, but with a frequency given by the difference between the respective Rabi frequencies, or a much lower frequency than with field gradient.). Alternative approaches that could be used to produce a local static magnetic field gradient without Joule heating are a micro-magnet[4,10] or superconducting on-chip antenna.

In conclusion, our measurements show for the first time photon-assisted tunneling in gate-defined Si/SiGe quantum dots. Charge stability of the device is only mildly affected. This demonstrates the feasibility of applying microwaves in a Si/SiGe double quantum dot for performing electron spin resonance. We also explore the use of a d.c. current applied to the antenna for generating local, tunable magnetic field gradients. A field gradient around $1\,\mu$T/nm is achievable with a 2 mA d.c. current through the antenna, limited by Joule heating.

We thank F. Braakman, M. Shafiei, P. Barthelemy and T. Baart for discussions, Raymond Schouten for technical assistance, and the Dutch Foundation for Fundamental Research on Matter (FOM) and the European Research Council (ERC) for financial support. E.K. is a recipient of a fellowship from the Nakajima Foundation. J.R.P., D.E.S., M.G.L., and M.A.E. acknowledge support from the U.S. Army Research Office (W911NF-12-1-0607, W911NF-08-1-0482). Development and maintenance of the growth facilities used for fabricating samples is supported by DOE, Grant # DE-FG02-03ER46028.

Figure 1

(a) Scanning electron micrograph of a device with identical design to the one we used. The quantum dot is formed at the locations of the left or right circle, depending on the measurement. Transport measurements are performed by applying a voltage between the source and drain reservoirs (S and D) and monitoring the current $I_{dot}$ through the dot. The microwave antenna, on the right of the image, consists of a short wire connecting the two arms of a coplanar stripline.

(b) Measured current through the left dot as a function of the voltage on gate 4, under microwave irradiation via the on-chip antenna at $f = 20\,\text{GHz}$ (red lines; the microwave source emits +10 dBm, there is a -10 dB attenuator at room temperature, and a -20 dB attenuator at 1 K) and in the absence of the microwave radiation (blue lines). $V_{SD} = -50\,\mu\text{V}$ in both cases (all $V_{SD}$ include thermal voltages).

(c) Histogram of the charge noise expressed in units of gate voltage with microwave excitation (red) and without microwave excitation (blue).

Figure 2

(a) Measured current through the right dot as a function of the voltage on gate 2 for different microwave powers and frequencies applied to the antenna. $V_{SD} = -58\,\mu\text{V}$ (the lines are offset for clarity). The 10 mV shift in the Coulomb peak position between 16.5 GHz and 20 GHz is due to a background charge switch, which occasionally occurs in this sample, both with and without microwave excitation.

(b-e) Schematics of the energy diagram of the quantum dot for $V_{SD} < 0$. (b,c) shows PAT through the right barrier at two different gate voltages. (d,e) show PAT through the left barrier at two different gate voltage. The net electron flow is from the drain to the source in (b,e) and from the source to the drain in (c,d).

Figure 3

(a-c) Measured current through the right dot as a function of the voltage on gate 2 with different d.c. currents through the antenna (see symbols in the inset of (c)). The voltage difference between the source and drain is $V_{SD} = -58 \mu V$ in (a), $-8 \mu V$ in (b) and $42 \mu V$ in (c). The solid lines are fits to Eq.(4) with the temperatures in the source $T_S$ and in the drain $T_D$ as fitting parameters. Insets in (a-c) show schematics of the energy levels for the corresponding $V_{SD}$, and for the case where the temperature is higher in the source reservoir than in the drain.

(d) Temperatures in the source and drain reservoir as a function of d.c. current through the antenna extracted from the fits in panels (a-c). The error bars indicate 95 % confidence intervals for the fitting parameters $T_S$ and $T_D$.

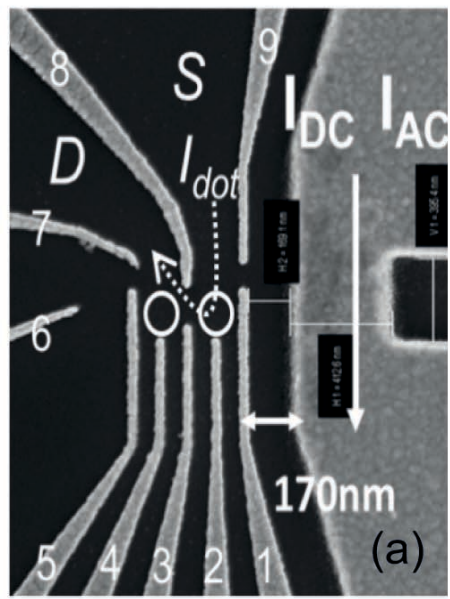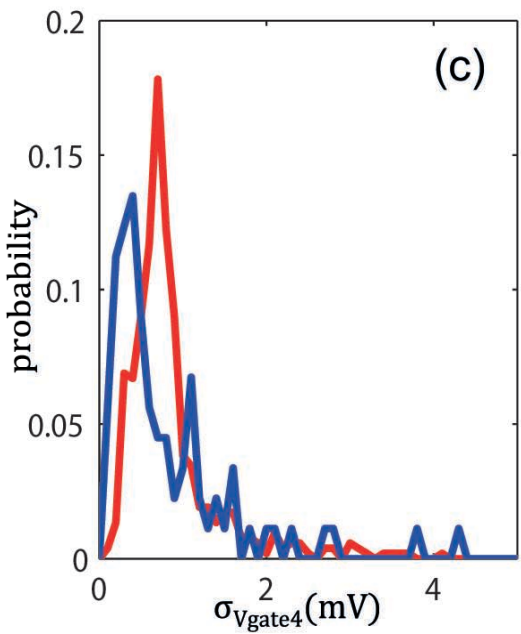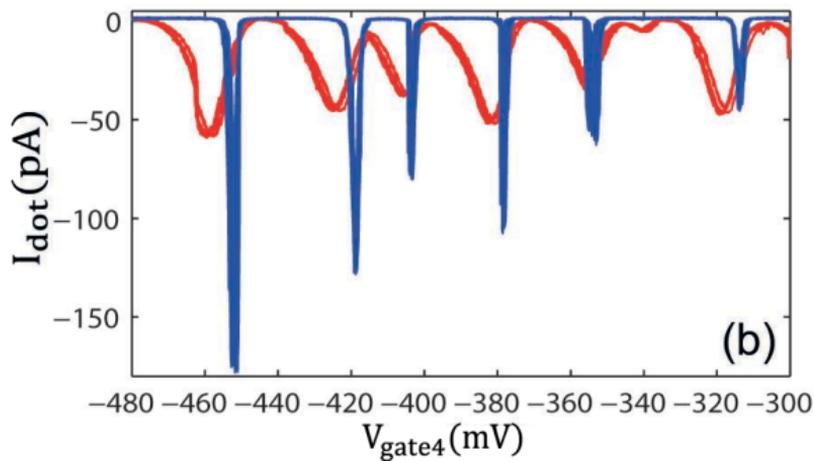

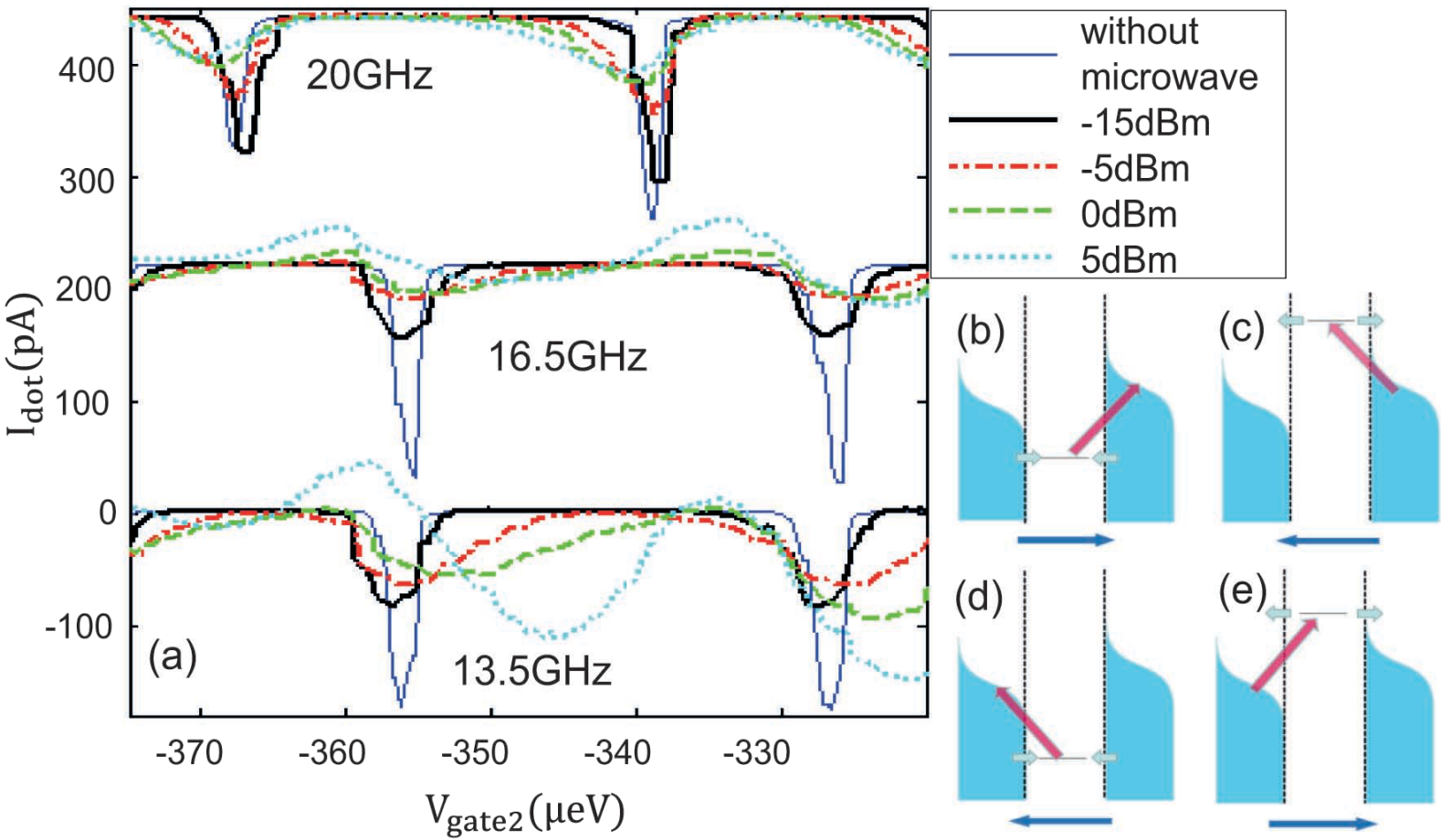

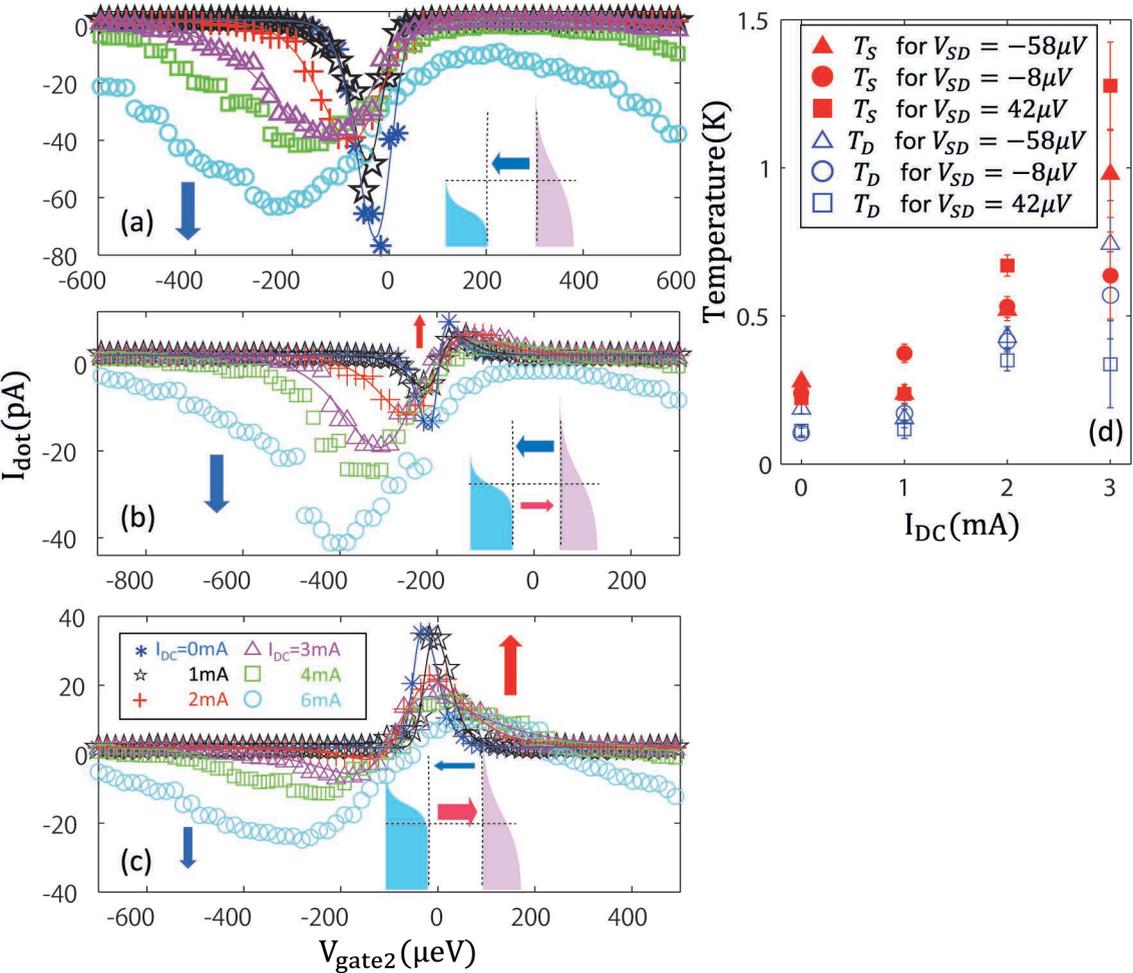